\documentclass[aps,preprint,epsf,prl]{revtex4}
\begin{document}
\title{Interaction of vacancies with grain boundary in aluminum: 
a first-principles study} 
\author{Gang Lu and Nicholas Kioussis}
\address{Department of Physics, California State University Northridge,
Northridge, CA 91330-8268}
\begin{abstract} 
We present a theoretical study of the interaction of vacancies with a tilt 
grain boundary in aluminum based on the density functional theory. 
The grain boundary volume expansion and vacancy induced contraction are 
calculated and compared for the nearest-neighbor atoms, and the former is found 
to be smaller than the latter. The formation energy of a vacancy placed at 
various layers in the grain boundary has been calculated and we find that 
the grain 
boundary does not always act as sinks for vacancies. In fact, it costs more 
energy to form a vacancy at the boundary plane than in bulk, although the rest 
of the grain boundary region does attract vacancies. The microscopic mechanisms 
of grain boundary sliding and migration are investigated thoroughly with and 
without a vacancy. We find that although the vacancy can hinder the grain 
boundary motion by tripling the energy barrier of sliding and migration, it can 
not inhibit or even delay the migration process. The vacancy placed at the first 
layer from the interface is found to be trapped at the layer and 
not able to follow
 the migrating interface.

\end{abstract}
\maketitle 
\begin{center}
I. INTRODUCTION
\end{center}

Vacancies are the simplest lattice defects and play an important role for 
various properties of materials, such as kinetic, thermodynamic, electrical, 
optical and even mechanical properties, to name a few. Similarly, grain 
boundaries as planar defects, are also responsible for the various material 
properties.\cite{wolf,sutton} While there have been extensive theoretical 
studies aimed at either vacancies or grain boundaries individually, much less 
effort has been devoted to the understanding of the collective behavior of grain 
boundaries and vacancies, especially from parameter-free {\it ab initio} 
studies. For example, {\it ab initio} calculations have been used with great 
success to study the vacancy formation energy and the vacancy-induced relaxation 
in different metals,\cite{turner,meyer} the vacancy diffusion energy barrier 
and the annihilation process in semiconductors,\cite{watanabe,pankratov,fazzio} 
etc. However, all these calculations are confined to either bulk systems or 
surfaces. On the other hand, grain boundaries (pure or with impurities) have 
attracted a great deal of attention recently. For example, {\it ab initio} 
calculations have been performed for the grain boundary energies and the atomic 
structures of the pure $\Sigma$11 tilt and $\Sigma$3 twin grain boundaries in 
aluminum\cite{wright}, and the effect of impurity on the grain boundary 
cohesion in Ni$_3$Al.\cite{lu} However, these studies do not involve vacancies 
in the grain boundaries and examine only the properties of the grain boundaries 
themselves. Therefore, one of the goals in this paper is to investigate the 
combined effect of grain boundaries and vacancies in aluminum using {\it ab 
initio} calculations and to explore the 
possible interactions between the point and planar defects.

While the equilibrium structures of grain boundaries have been extensively 
studied, only very limited research has focused on the atomistic simulation of 
grain boundary sliding and migration.\cite{mishin,hoagland,chandra,bishop}  
The former process involves the 
relative displacement of the two constituent grains in a direction parallel to 
the boundary interface, while the latter involves the motion of the interface 
in a direction perpendicular to the boundary plane. Grain boundary 
sliding/migration is considered to be one of the principal mechanisms of plastic 
deformation of polycrystalline materials at intermediate-to-high temperatures 
(above 0.6T$_m$, where T$_m$ is the melting point) and it contributes to creep 
deformation and intergranular fracture, which can in turn be significantly 
affected by the presence of vacancies.\cite{sutton,nabarro} These two modes of 
grain boundary motion can be correlated from a simple geometrical analysis. 
\cite{ashby} Very recently, Molteni {\it et al.} carried out {\it ab initio} 
studies of the atomic structure and the energy barriers for grain boundary 
sliding in germanium and aluminum.\cite{molteni1,molteni2} 
Employing molecular dynamics simulations in conjunction with the Embedded Atom 
Method, Chandra and Dang have shown that the coupling between sliding and 
migration depends strongly on the external conditions that initiate the grain 
boundary sliding.\cite{chandra}

In this paper we have studied the interactions of vacancies with the tilt 
$\Sigma$5 (210)[001] grain boundary in aluminum using state-of-the-art {\it 
ab initio} total-energy calculations based on the pseudopotential plane wave 
method. 
First, we have calculated the vacancy formation energies and atomic structures 
for a vacancy placed at various positions in the grain boundary region. These 
results 
reveal that the grain boundary may act as sink or source of vacancies 
depending on the location of the vacancy. Second, in order to understand the 
effect of vacancies on the grain boundary mobility, we have investigated the 
effect of vacancies on the grain boundary sliding and migration behavior.
Aluminum is selected in this study because it is a prototypical simple metal of 
importance in industrial applications and some of the conclusions we draw from 
the present study may be extended to other fcc metals as well. 

The remainder of this paper is organized as follows: Sec. II describes briefly 
the computational techniques and the supercell used for the grain boundary 
calculations. In Sec. III we present the results for the atomic structures and 
energetics of the pure grain boundary. Sec. IV contains the results with 
vacancies introduced at various atomic sites of the grain boundary. The vacancy 
formation energy at different sites is calculated and compared to that of 
the bulk system. The atomic relaxation due to the grain boundary expansion and 
the vacancy-induced contraction will be examined and compared. 
In Sec. V we present the energy profile and the atomic structures for the grain 
boundary sliding and migration process with and without a vacancy. We find that 
the vacancy can significantly increase the grain boundary sliding and migration 
energy barrier, and therefore hinder the grain boundary motion. In order to shed 
light into the microscopic physics of the combined grain boundary 
sliding/migration motion, we have examined in detail some critical atomic 
configurations during the sliding/migration process. In Sec. VI a brief summary 
and statement of conclusions are presented. 

\begin{center}
II. MODEL AND COMPUTATION
\end{center}

The electronic structure of the grain boundary is calculated by means of the 
pseudopotential plane wave method based on the density functional theory and the 
local density approximation. We employed the exchange and correlation potential 
of Ceperly and Alder as parametrized by Perdew and Zunger\cite{perdew} and the 
norm-conserving nonlocal pseudopotential  of Bachelet, Hamann and Schl\"{u}ter.
\cite{BHS} The Kohn-Sham wave functions are expanded in plane waves with a 
kinetic energy cutoff of 14 Ry. The atomic structures are considered fully 
relaxed when the Hellmann-Feynman forces on each atom are smaller than 0.001 
Ry/au. For the Brillouin-zone sampling, a $k$-point grid consisting of (2,7,1) 
divisions along the reciprocal-lattice directions is used according to the 
Monkhorst-Pack scheme.\cite{monk} Convergence tests have been performed both for 
the number of $k$-points as well as the number of plane waves. The calculated 
equilibrium lattice constant ($a_0$), bulk modulus and cohesive energy are 3.93 
\AA, 80.59 GPa and 3.26 eV/atom, respectively, 
in good agreement with the corresponding 
experimental room-temperature values of 4.05 \AA, 76.93 GPa and 3.39 eV/atom.  

In this work we consider the symmetric $\Sigma$5 (210) [001] tilt grain boundary 
in the coincident-site-lattice (CSL) notation. In the CSL notation, the grain 
boundary is formed through rotation of the two fcc grains 
about the [001] tilt axis by
36.9$^\circ$. The boundary is oriented along the (210) plane and the inverse 
density of CSL sites, $\Sigma$ = 5, characterizes the unit cell volume of the 
CSL superlattice. The $\Sigma$5 [001](210) grain boundary of Al is simulated by 
the orthorhombic supercell shown in Fig. 1. The supercell contains 60 atoms 
distributed in 30 (210) layers parallel to the interface. The large and small 
circles represent the atoms on the (001) and (002) planes, respectively. The 
dimensions of the supercell along the [\={1}20], [001] and [210] directions are 
8.78 \AA $\times$ 3.93 \AA $\times$ 26.35 \AA, respectively. 
Because of the periodic boundary conditions, the supercell contains two 
equivalent grain boundaries. However, as will be shown in Sec. III, the 
separation between the two grain boundaries is large enough to eliminate the 
interaction between them. For the modeling of vacancies in the grain 
boundary, a single vacancy is placed at all possible independent lattice sites 
labeled by an integer {\it n} = 0-7, which represents the {\it n}th 
nearest-neighbor layer from the grain boundary plane (the number of independent lattice 
sites is reduced due to the symmetry between the (001) and (002) planes and that 
between the two grain boundaries). In addition to the grain boundary 
calculation, we have also computed the vacancy formation energy for the Al bulk 
by employing a simple cubic supercell with 108 atomic sites to model an isolated 
vacancy.

\begin{center}
III. PURE GRAIN BOUNDARY
\end{center}

A pure grain boundary in the following is referred to as the grain boundary 
without vacancies. The atomic structure of the pure grain boundary before and 
after relaxation is shown in Fig. 2. The open and filled circles represent the 
atomic positions before and after the relaxation and the large and small circles 
correspond to the atoms on the (001) and (002) planes, respectively.  
The grain boundary volume expansion is clearly seen near the interface and one
local measure for such an 
expansion is the relative normal displacement of the 
two atomic planes closest to the interface. The present calculation yields a 
local expansion of 0.09 $a_0$, where $a_0$ is the bulk Al 
lattice constant. This local expansion of the Al atoms in the pure Al $\Sigma$5
grain boundary is smaller compared to the corresponding expansion of 
the Al atoms in the Ni$_3$Al alloy for the same grain boundary.\cite{lu}
This is consistent with the fact that there is charge transfer from the Al atoms 
to the nearest-neighbor Ni atoms in Ni$_3$Al, which in turn reduces the 
screening charge between the nearest Al-Al ions across the interface, leading to 
a stronger electrostatic repulsion between the Al ions in Ni$_3$Al grain 
boundary. In Fig. 3 we present the relative variation (strain) of the interlayer 
spacing as a function of the layer away from the grain boundary plane. 
The strain has a symmetric oscillatory profile that reaches its maximum at the 
first layer and decays into the bulk. The nature of this strain profile can be 
traced to the Friedel oscillations of the electron distribution which drive the 
ions to relax in such an oscillatory pattern.\cite{cho}  The negligible strain 
for the layers from 6 to 10 clearly indicates that the atomic relaxation is 
localized only within 5 layers from the interface. Therefore, the supercell with 
15 layers between the two interfaces is large enough to capture the overall 
properties of the grain boundary. 

It is interesting to notice that even after the relaxation, the distance between 
the nearest-neighbor atoms across the interface is 2.44 \AA, about 20\% smaller 
than the corresponding distance of 2.78 \AA~  in the bulk system. 
This result invalidates 
the common misconception that the grain boundary is associated with open space. 
On the other hand, as will be shown in Sec. IV, vacancies can be formed 
relatively easier in most of the grain boundary region compared to the bulk, 
giving rise to more empty space in the region. The grain boundary energy can be determined 
from the difference of the energy of a
 unit cell containing the grain boundary and a 
unit cell containing an equal number of atoms in the bulk environment, divided 
by the grain boundary area. We find a grain boundary energy of 502 mJ/m$^2$ for 
the $\Sigma$5 grain boundary, which is larger than 
the corresponding EAM value of 
351 mJ/m$^2$.\cite{chen} This result is consistent with the finding that, overall,  the EAM 
potential tends to give smaller grain boundary energies compared to {\it ab 
initio} calculations.\cite{wright}

\begin{center}
IV. GRAIN BOUNDARY WITH A VACANCY
\end{center}

The relaxed atomic structures of the grain boundary before and after the 
introduction of a vacancy are shown in Fig 4.(a), (b), (c), corresponding to the 
vacancy placed at the zero-th (interface), first and seventh layers from the 
interface. The large and small circles correspond to atoms on the (001) and 
(002) planes, respectively. The open (filled) circles represent the relaxed 
atomic positions before (after) the introduction of the vacancy, and the arrows 
show the direction of the atomic displacements. In order to gain insight into 
the interplay between the grain boundary and the vacancies, we have calculated 
the displacements due to the grain boundary expansion (u$_{GB}$) and that due to
the vacancy contraction (u$_{V}$). Both u$_{GB}$ and u$_{V}$ are computed for 
the nearest-neighbor atoms (first shell) from the defects. For the case 
of the vacancy at the interface (Fig. 4(a)), although the atomic 
relaxation is small and localized, 
the vacancy-induced contraction (u$_{V}$ = -0.12 \AA) is much larger than 
the grain boundary expansion (u$_{GB}$ = +0.01 \AA). When the vacancy is placed 
at the first layer, the atomic relaxation is more dramatic (Fig. 4(b)). 
There are seven atomic layers involved in the relaxation triggered by the 
vacancy and the first shell displacements are u$_{V}$ = -0.53 \AA ~
while u$_{GB}$= 
+0.34 \AA, respectively. In Fig. 4(c) we show the atomic structure with the 
vacancy placed at the seventh layer (bulk like). We find u$_{V}$ = -0.06 \AA, 
which is about 2\% of the nearest-neighbor distance of the bulk lattice and is 
along the [110] direction. These results are in excellent agreement with the 
calculations by Turner {\it et al.} \cite{turner} for a vacancy in Al bulk both 
for the magnitude and the direction of u$_{V}$. In other words, the atomic 
structure for a vacancy at the seventh layer from the interface completely 
reproduces the characteristics of a vacancy in the bulk. Along with the fact 
that the grain boundary expansion u$_{GB}$ at the seventh layer is zero as  
discussed earlier, we conclude that the selected supercell for the grain 
boundary calculations is large enough to yield reliable results. It is also 
interesting to note that as far as the nearest-neighbor shell is concerned, the 
vacancy-induced atomic relaxation is more pronounced than that due to the grain 
boundary. 

In order to clarify whether grain boundaries act as sinks or 
sources for vacancies, 
we have calculated the vacancy formation energy 
for various layers from the grain 
boundary interface. The results are presented in Fig. 5, where the straight line
denotes 
the vacancy formation energy in bulk. The striking feature of the energy curve 
is the large reduction 
of the vacancy formation energy at the first layer, which strongly suggests the 
preference of forming a vacancy at the first layer compared 
to bulk. Although the 
figure also shows that some other more distant layers in the grain boundary are 
also energetically favorable for the vacancy formation, one can clearly see that 
the interface itself is {\it not} favorable for forming a vacancy. In fact, 
there is more than 0.1 eV energy penalty for creating a vacancy at the interface 
compared to bulk. 
Note that the vacancy formation energy at the fifth layer from the interface 
is close to the bulk value,
 showing the convergence of the energy with respect to the supercell size.
In conclusion, the {\it ab initio} calculations support the traditional view 
that grain boundaries act as pathways for vacancy diffusion and that vacancies 
prefer to segregate to grain boundaries over bulk at low stress.  However, we 
should also point out that one out of five lattice sites in the grain boundary 
region turns out to be energetically unfavorable for vacancies, namely those 
sites on the grain boundary plane.

\begin{center}
V. GRAIN BOUNDARY SLIDING AND MIGRATION: EFFECT OF VACANCIES
\end{center}

In this section we present results for the grain boundary sliding and migration, 
and the effect of vacancies on the grain boundary motion.

In our {\it ab initio} modeling, the grain boundary sliding is simulated
quasi-statically, namely the top grain is rigidly shifted over the bottom by a series 
of small specified distances along the interface. The sliding distance
is described in 
percentage of $a_{csl}$, where $a_{csl}$ is the lattice parameter of the CSL 
cell along [\={1}20]. Clearly sliding by $a_{csl}$ will bring the 
grain boundary back to its original configuration due to the imposed 
periodic boundary 
conditions. For each rigid shift, atomic relaxation is performed to locate the 
closest local energy minimum configuration. In order to understand the effect of 
vacancies on the grain boundary sliding and migration process, we have 
calculated the grain boundary energy profile as a function of sliding distance 
for the pure grain boundary and the grain boundary with a vacancy placed at the 
first layer from the interface, corresponding to the lowest vacancy 
formation energy at the grain boundary. 

The grain boundary energy profile during the sliding process is shown in Fig. 6, 
where the filled circles represent the pure grain boundary and the open circles 
represent the grain boundary with the vacancy, respectively. The labels (a)-(f) 
 on both curves indicate the special atomic configurations which will be 
explored in detail below. For the pure grain boundary, there are two energy 
peaks at 30\% and 80\% $a_{csl}$, the first much broader than the latter; 
and one energy 
valley at 55\% $a_{csl}$.
 It is important to note that the vacancy triples the energy 
barrier for sliding and migration, which demonstrates the great hindering effect 
of the vacancy on the interface mobility. However, the detailed analysis of the 
atomic structures below shows that the vacancy can not prevent or even delay the 
occurrence of the grain boundary migration. 

The collective motion of atoms during the grain boundary sliding and migration is 
essential to understand this seemingly trivial, but in fact complicated process. 
To shed light on the related physical process, we next examine the atomic 
configurations associated with the special points (a)-(f) in the energy 
curves of the pure grain boundary 
in Fig. 6. The relaxed atomic structures corresponding to 
(a)-(f) are shown clockwise in 
Fig. 7, with the filled and open circles representing the atoms on the (001) and 
(002) planes, respectively. Note that more atoms are displayed in these figures 
than the actual atoms used in the supercell, 
which is depicted by the solid frame. In 
order to monitor when does grain boundary migration take place, one needs to 
locate the position of the interface during the sliding process. However, 
since the grain boundary is not symmetric any more during the sliding process, 
it is not always obvious to find the new position of the migrating interface. 
To overcome this difficulty, we propose a simple scheme which allows us to 
identify the moving interface by following the 
evolution of the structural unit shown in the 
figures. The characteristics of this unit is that it consists of two 
sub-units that cross the interface with a common vertex on the interface. 
These sub-units are the two-dimensional unit cells of the distorted fcc lattice. 
By locating such a unit and the common vertex during the sliding process, one 
can identify the position of the grain boundary plane. Applying this scheme, we 
find that the migration of the pure grain boundary starts at 35\% $a_{csl}$ and 
ends at 55\% $a_{csl}$. After the completion of the first migration, the 
grain boundary has to return to its original position at 100\% $a_{csl}$ to 
complete the second migration.
For the grain boundary with the vacancy, the migration starts at the exact 
same sliding distance, 35\% $a_{csl}$, but ends at about 62\% $a_{csl}$. The 
second migration due to the periodic boundary conditions is also observed as in 
the pure grain boundary.

Fig. 7(a) shows the initial atomic structure for the pure grain boundary before 
sliding. The interface is indicated by a solid line. 
The angle $\alpha$ ($\angle$ABC) of the structural unit is slightly distorted 
from its value of 90$^\circ$ in the perfect fcc lattice. By examining the various atomic 
configurations before the first migration (35\% a$_{csl}$), we find that they 
are all similar, being different mostly along the sliding direction. More 
specifically, the angle $\alpha$ of all configurations turns 
out to be obtuse and 
one example of such a configuration is shown in Fig. 7(b). When the interface 
starts migrating, however, the grain boundary experiences an abrupt overall 
structural transformation, as illustrated in Fig. 7(c).  The interface has 
clearly moved from its original position (dotted line) to the current solid line 
position, and the angle $\alpha$ has changed from obtuse to acute. This 
one-layer migration reflects the geometrical necessity for the coupling between 
migration and sliding.\cite{bishop,ashby} After the onset of the migration, the 
atomic structures remains akin to each other during the sliding until the 
migration is fully completed, in which, as shown in Fig. 7(d), the grain 
boundary recovers its original symmetric structure. Note, that the angle 
$\alpha$ now changes from an acute angle to a right angle. The dotted and solid 
lines are the same as in Fig. 7(c). Because of the periodic boundary conditions, 
the grain boundary has to return to its original position during the remaining 
sliding process. The critical configuration it has to cross is shown in 
Fig. 7(e), where the grain boundary experiences the highest energy. The 
intriguing feature of this configuration is that no boundary plane can be 
identified, while there is a ``mirror'' plane (dashed line) across which the 
atoms on the (001) plane are symmetrically situated to those on the (002) plane. 
This half-way plane reflects the transient position of the migrating interface 
and therefore the energy associated with this configuration gives the energy 
barrier for the grain boundary migration. Finally, Fig. 7(f) represents the 
completion of the sliding process where the grain boundary returns to its 
initial symmetric configuration. The dotted line indicates the position of the 
interface after the first migration and the solid line shows the current 
position (also the initial position before the sliding) of the interface. 

For the case of the vacancy placed at the first layer from the 
grain boundary, the sliding energy 
profile becomes more irregular, with multiple peaks and valleys (Fig. 6). The 
abundance of the fine grain boundary structures reflects the complexity of the 
system, so that we shall concentrate only on the most important events. Fig. 
8(a) shows the atomic structure before sliding and the vacancy is denoted by the 
letter V. The solid line indicates the current position of the interface. It is 
found that this initial state no longer corresponds 
to the lowest energy, which is 
reached after a sliding displacement of 2\% a$_{csl}$.  There is an energy 
cross-over at 10\% a$_{csl}$, above which the grain boundary energy with the 
vacancy becomes larger than the corresponding value of the pure grain boundary, 
until the next cross-over occurs at about 75\% a$_{csl}$. Although the energy 
value does not return to zero as in the case of the pure grain boundary when the 
migration is completed, careful examination of the atomic structures in Fig. 8 
indicates the occurrence of the grain boundary migration. 
Just like the pure grain 
boundary, the atomic structures remain similar to each other during the sliding 
process until the first migration takes place at about 35\%. One such 
configuration is shown in Fig. 8(b), where the angle $\alpha$ is obtuse. The 
highest energy configuration is reached when the interface migrates one atomic 
layer upward, shown in Fig. 8(c), in which the dotted and solid lines represent 
the original and current positions of the interface, respectively. As in 
the case of the pure grain boundary, the 
atomic structures at the onset of the migration differ significantly from those 
before the migration and the atoms near the interface become severely distorted 
from their positions in the underlying fcc lattice. The grain boundary completes 
its first migration at about 62\% a$_{csl}$, where the symmetric structure is 
recovered (Fig. 8(d)). It is important to notice, however, that this symmetric 
structure with the vacancy (labeled by V) at {\it the interface}
 is different from the 
initial state where the vacancy was at 
{\it the first layer} above the interface. 
Therefore, the energy difference between the initial state (a) and the current 
state (d) reflects simply the difference in the vacancy formation energy 
between the vacancy placed at the 
interface and the first layer, respectively. 
This indicates that the vacancy is trapped in the 
first layer and not able to follow
 the migrating interface. The grain 
boundary develops another peculiar configuration at 92\% a$_{csl}$, just before 
the completion of the second migration. A symmetric pattern forms relative 
to the previous interface (dashed line) and the free volume associated with the 
vacancy disappears. The off-stoichiometry due to the vacancy, however, is 
recovered at the symmetry plane (dashed line). The solid line represents the 
final (also initial) position of the interface during the entire sliding 
process. The atomic structure of the grain boundary at 100\% a$_{csl}$ 
is shown in Fig. 8(f) with the letter V indicating the vacancy. The dotted 
(solid) lines represent the grain boundary position after the first (second) 
migration.

\begin{center}
VI. CONCLUSION
\end{center}

We have studied the interaction between vacancies and the $\Sigma$5 tilt grain 
boundary in Al, using {\it ab initio} pseudopotential calculations, which  
yield reliable results for both the atomic structure and energetics compared 
to empirical atomistic approaches. The nearest-neighbor distance across the interface is actually 
smaller than that in the bulk even after atomic relaxation. We have calculated 
the grain boundary volume expansion and the vacancy-induced 
contraction and found 
that the latter is larger 
than the former for the nearest-neighbor shell. We have also calculated the vacancy formation energy 
for a vacancy placed at different layers from the interface. We find that 
the grain boundary does not always act as sinks for vacancies. While most of the 
grain boundary sites have lower vacancy formation energy compared to bulk, those 
at the interface have higher formation energy. We have investigated the combined 
grain boundary sliding and migration process with and without a vacancy. We have 
introduced a simple scheme to identify the migrating interface during the 
sliding process. We have found that although the vacancy can hinder the grain 
boundary motion by tripling the energy barrier for sliding and migration, it can 
not inhibit or delay the occurrence of migration. In fact, the grain boundary 
starts the migration at 35\% a$_{csl}$ regardless if there is a vacancy 
or not. We have 
analyzed the 
atomic configurations during the 
sliding process in order to gain 
better insight on this complicated process. Interestingly, the vacancy 
is found to be 
trapped at the first layer and not able to follow
 the migrating interface. 

\begin{center}
ACKNOWLEDGMENTS
\end{center}

The research was supported through the US Army Research Office under Grant No. 
DAAG55-97-1-0093.

\newpage
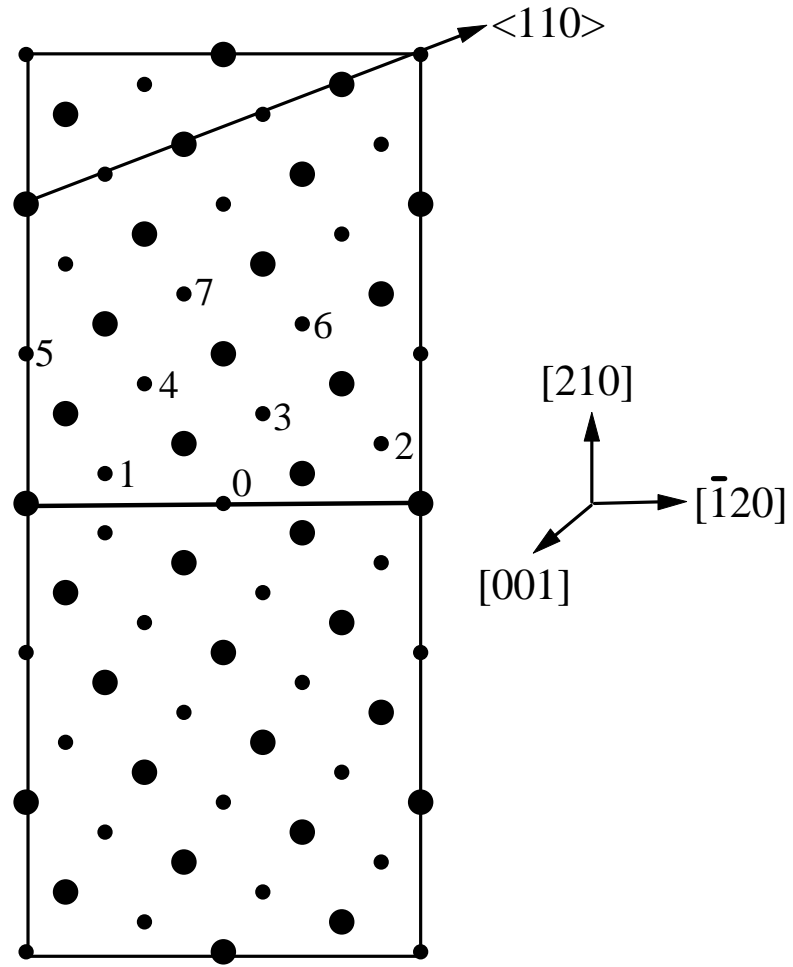
\begin{figure}
\caption{Supercell for the $\Sigma$5 (210) [001] tilt grain boundary in Al 
viewed along the [001] direction. The large and small circles represent atoms on 
the (001) and (002) planes, respectively. The integer (n = 0-7) represents the 
nth nearest-neighbor layer from the grain boundary interface
that a vacancy is placed. The grain boundary plane is indicated by the solid line.}
\label{fig1}
\end{figure}

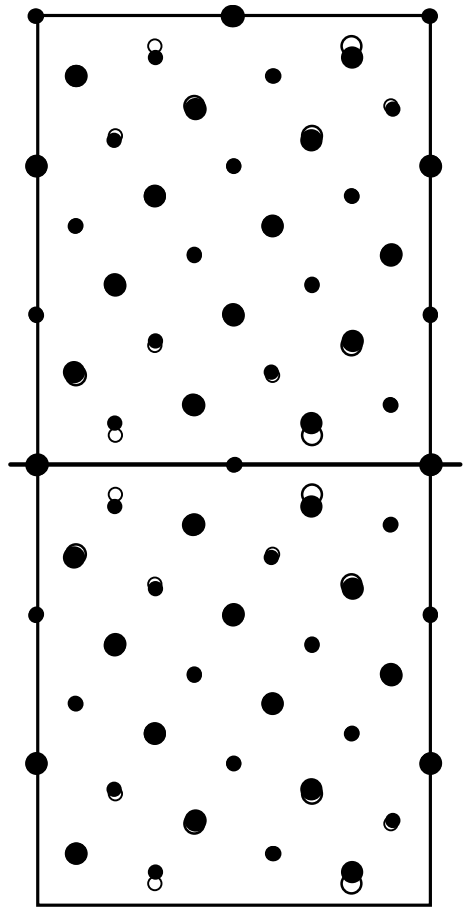
\begin{figure}
\caption{Unrelaxed (open circles) and relaxed(filled circles) atomic structure 
for the pure $\Sigma$5 grain boundary. Large and small circles represent atoms 
on the (001) and (002) planes, respectively.}
\label{fig2}
\end{figure}

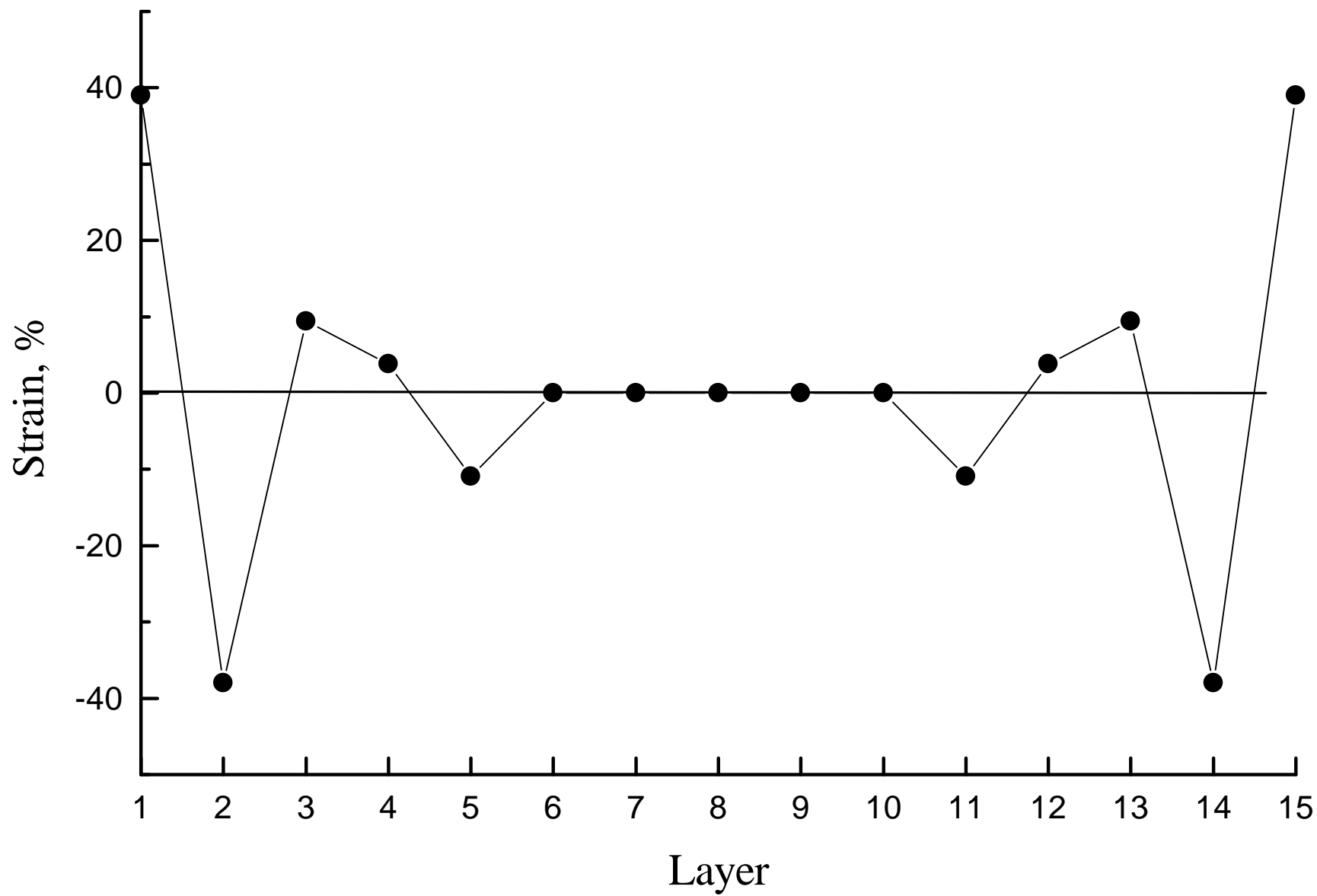
\begin{figure}
\caption{Interlayer strain (percentage) as a function of the number of layers 
away from the interface.} 
\label{fig3}
\end{figure}

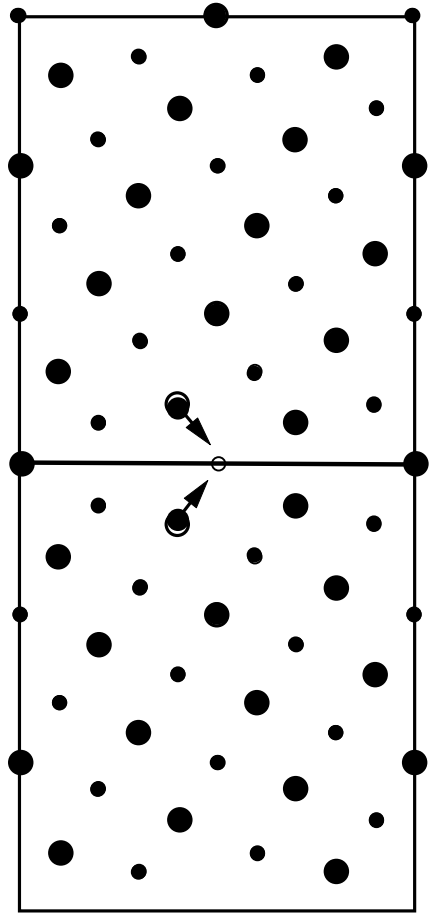
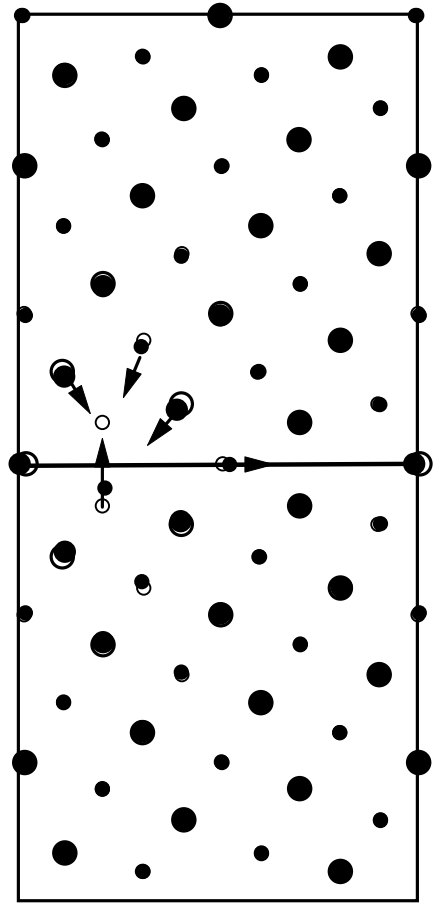
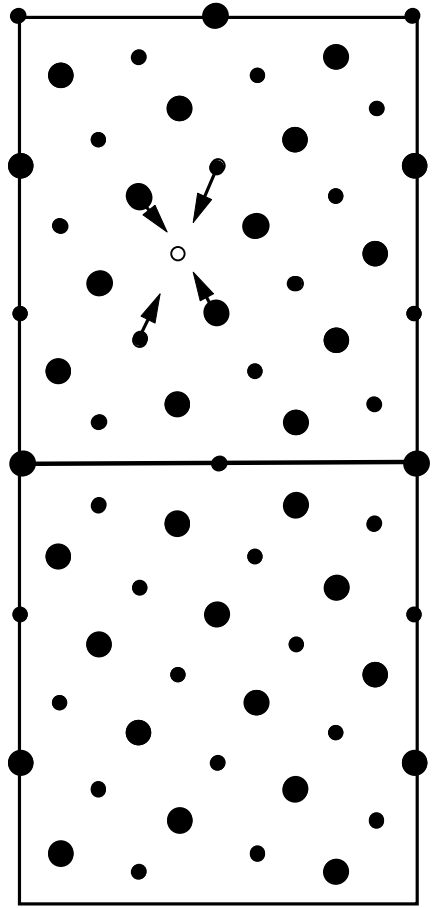
\begin{figure}
\caption{Relaxed atomic structures for the grain boundary before (open circles) 
and after (filled circles) the introduction of a vacancy at the interface (a), 
the first layer (b) and the seventh layer (c) from the interface, respectively. 
The large and small circles represent atoms on the (001) and (002) planes, 
respectively. The arrow shows the direction of the atomic displacement due to 
the vacancy.}
\label{fig4}
\end{figure}

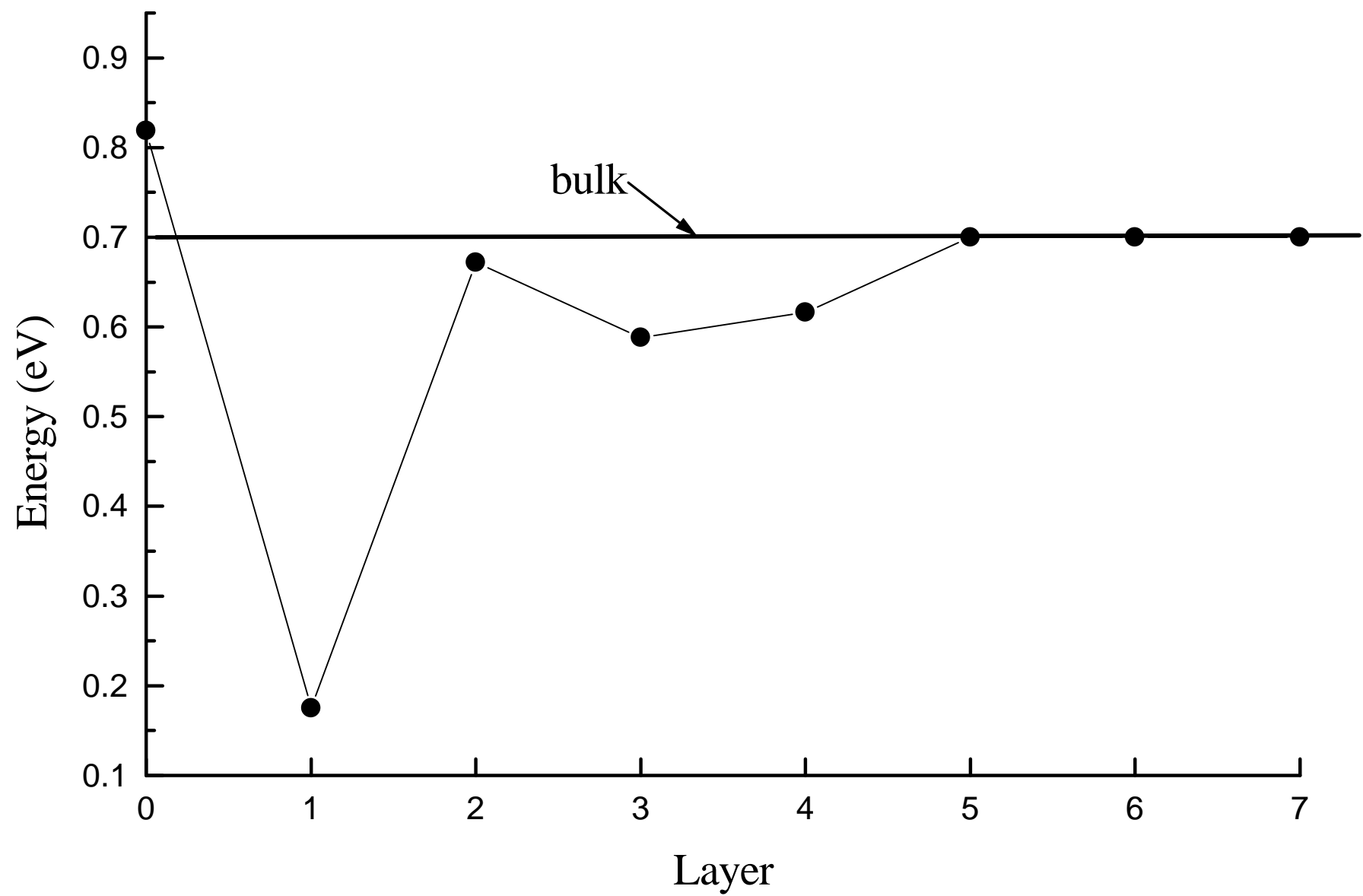
\begin{figure}
\caption{Vacancy formation energy as a function of the number of 
the layer away from the interface on which a vacancy is 
placed. The vacancy formation energy in bulk Al is denoted by the solid line.}
\label{fig5}
\end{figure}

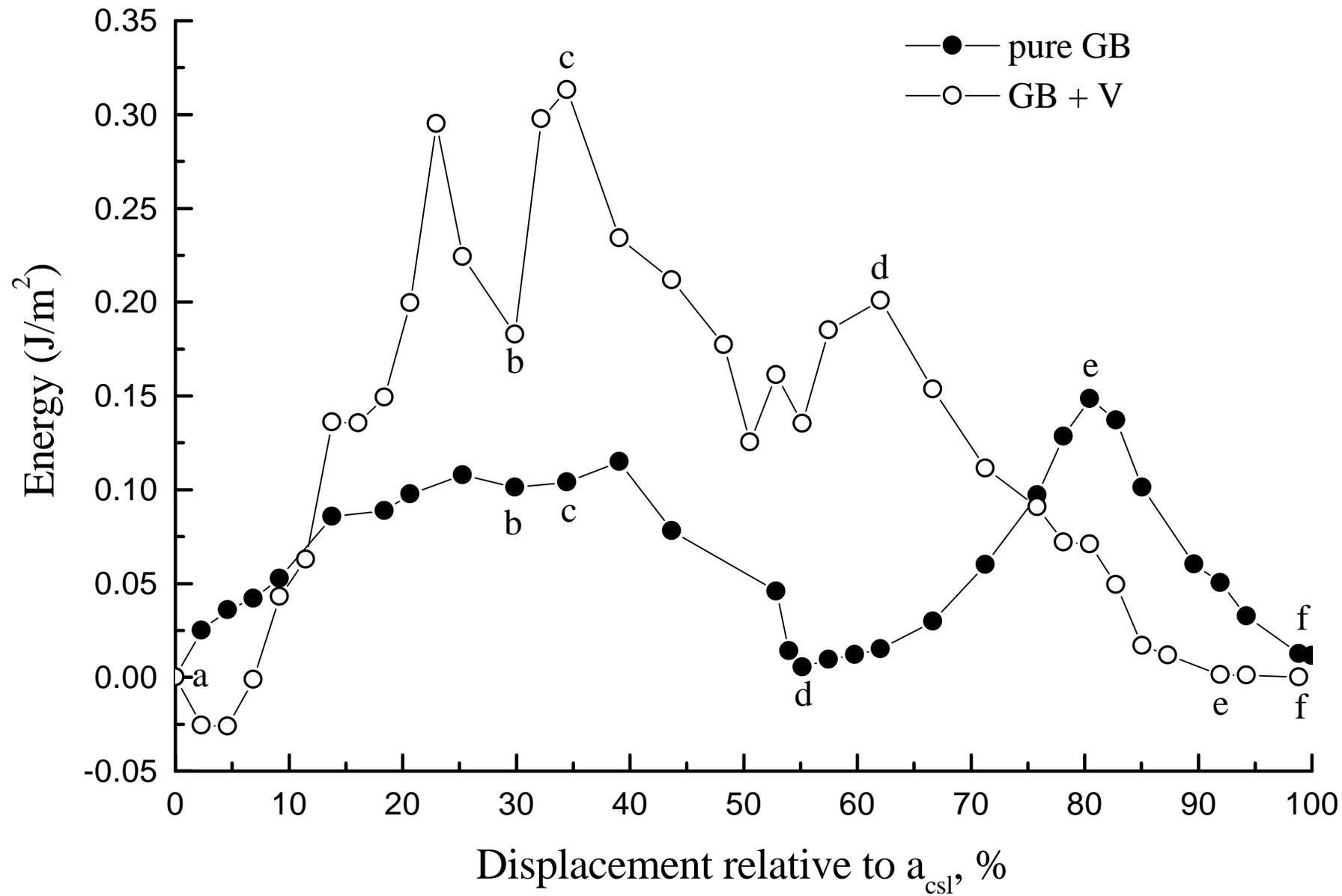
\begin{figure}
\caption{Relative grain boundary energy as a function of sliding distance 
expressed as percentage of the lattice parameter, a$_{csl}$, of the CSL cell 
along [\={1}20]. The filled and open circles correspond to the energies of the 
pure grain boundary and the grain boundary with a vacancy at the first layer, 
respectively. The labels from (a) to (f) represent the special atomic 
configurations displayed in Figs. 7 and 8.}
\label{fig6}
\end{figure}

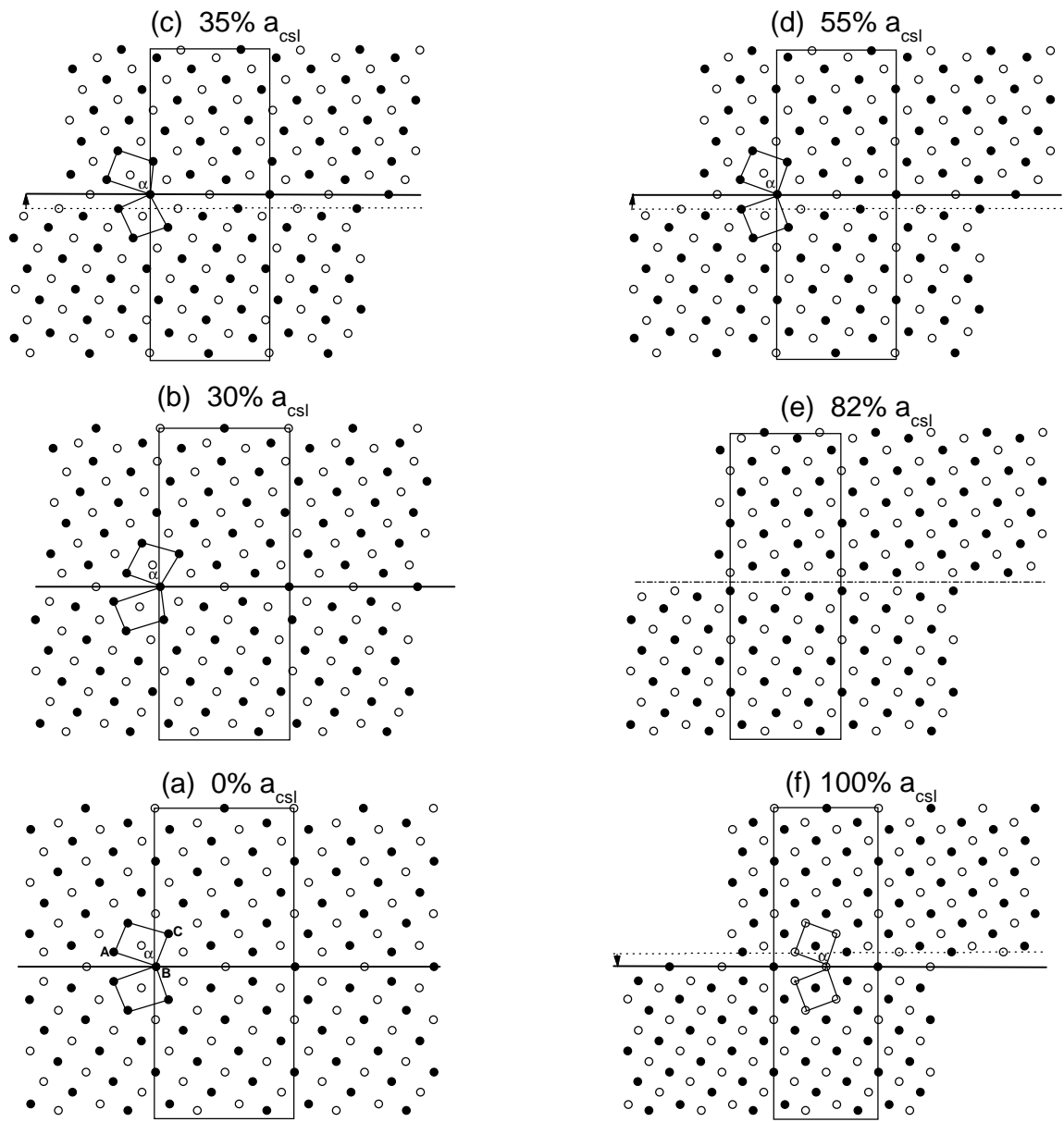
\begin{figure}
\caption{Evolution of the atomic structure of the pure grain boundary during the 
sliding process. The snap shots of the grain boundary structures correspond to 
the special points (a)-(f) on the energy curve for the pure 
grain boundary in Fig. 6. 
Filled and open circles represent atoms on the (001) and (002) 
planes, respectively, and the arrow indicates the direction of grain boundary 
migration.} 
\label{fig7}
\end{figure}

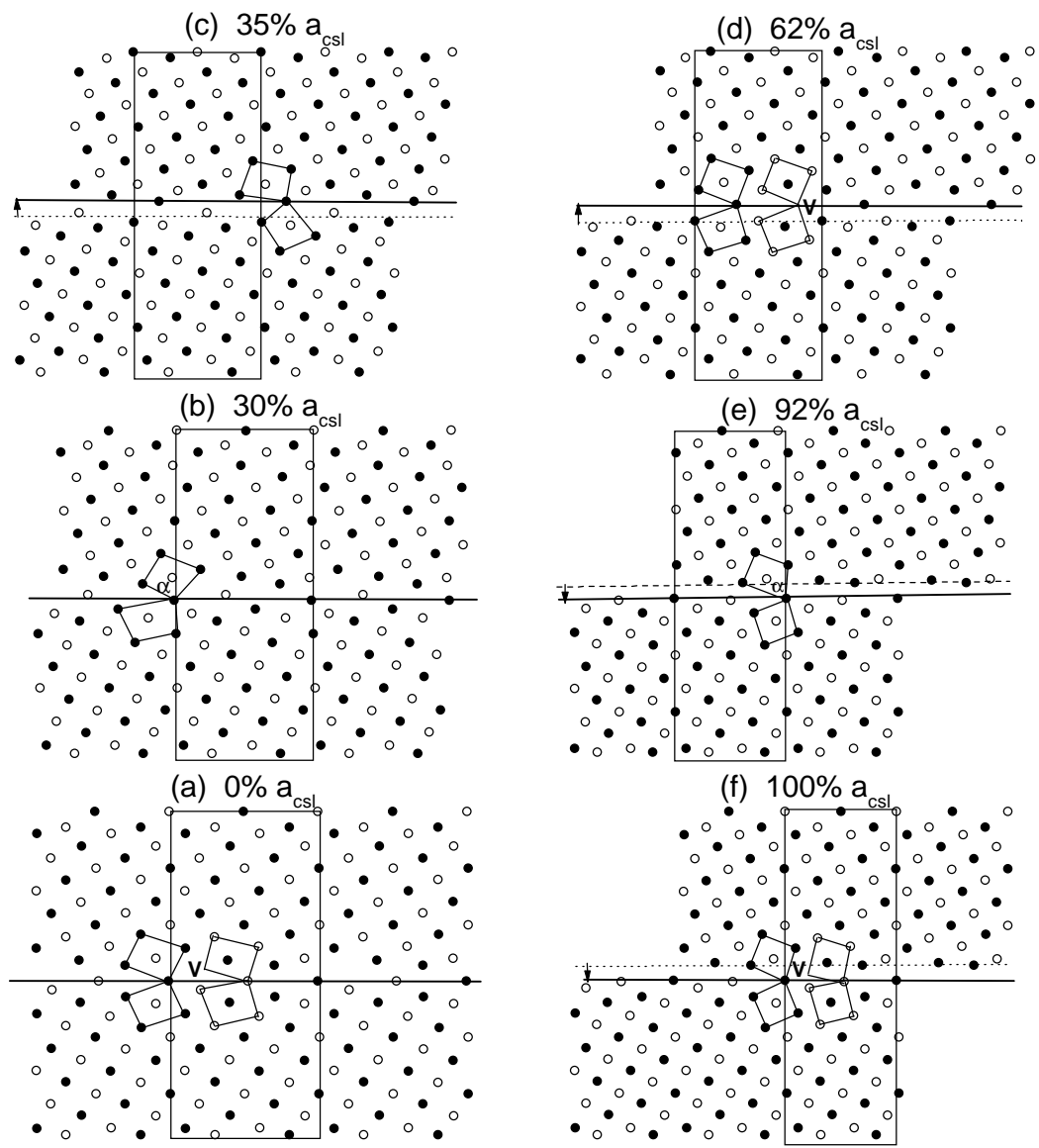
\begin{figure}
\caption{Evolution of the atomic structure of the grain boundary during the 
sliding process with a vacancy placed at the first layer above the interface. 
The snap shots of the grain boundary structures correspond to 
the special points (a)-(f) on the energy curve for the 
grain boundary with the vacancy in Fig. 6.
Filled and open circles represent atoms on (001) and (002) planes, respectively, 
and the arrow indicates the direction of grain boundary migration.}
\label{fig8}
\end{figure}

\newpage
\begin{thebibliography}{99}
\bibitem{wolf}
{\it Materials Interfaces - Atomic-level Structure and Properties}, edited by D. 
Wolf and S. Yip (Chapman and Hall, London, 1992). 
\bibitem{sutton}
{\it Interfaces in Crystalline Materials}, edited by A. P. Sutton and R. W. 
Balluffi (Oxford University Press, Oxford, 1995). 
\bibitem{turner}
D. E. Turner, Z. Z. Zhu, C. T. Chan, and K. M. Ho, Phys. Rev. B {\bf 55}, 13842 
(1997).
\bibitem{meyer}
B. Meyer and M. F\"{a}hnle, Phys. Rev. B {\bf 59}, 6072 (1999).  
\bibitem{watanabe}
H. Watanabe and M. Ichikawa, Phys. Rev. B {\bf 55}, 9699 (1997).
\bibitem{pankratov}
O. Pankratov, H. Huang, T. Diaz de la Rubia, and C. Mailhiot, Phys. Rev. B {\bf 
56}, 13172 (1997).
\bibitem{fazzio}
A. Fazzio, A. Janotti, A. J. R. da Silva and R. Mota, Phys. Rev. B {\bf 61}, 
R2401 (2000).
\bibitem{wright}
A. F. Wright and S. R. Atlas, Phys. Rev. B {\bf 50}, 15248 (1994).
\bibitem{lu}
G. Lu, N. Kioussis, R. Wu, and M. Ciftan, Phys. Rev. B {\bf 59}, 891 (1999).
\bibitem{mishin}
Y. Mishin and D. Farkas, Phil. Mag. A {\bf 78}, 29 (1998).
\bibitem{hoagland}
R. G. Hoagland and M. I. Baskes, Scripta Mater. {\bf 39}, 417 (1998). 
\bibitem{chandra}
N. Chandra and P. Dang, J. Mater. Sci {\bf 34}, 655 (1999).
\bibitem{bishop}
G. H. Bishop, R. Harrison, T. Kwok, and S. Yip, J. Appl. Phys. {\bf 53}, 5596 
(1982).
\bibitem{nabarro}
F. R. N. Nabarro and H. L. de Villiers, {\it The Physics of Creep}, (Taylor \& 
Francis, London, 1995).
\bibitem{ashby}
M. F. Ashby, Surf. Sci. {\bf 31}, 498 (1972).
\bibitem{molteni1}
C. Molteni, G. P. Francis, M. C. Payne, and V. Heine, Phys. Rev. Lett. {\bf 76}, 
1284 (1996).
\bibitem{molteni2}
C. Molteni, N. Marzari, M. C. Payne, and V. Heine, Phys. Rev. Lett. {\bf 79}, 
869 
(1997).
\bibitem{perdew}
J. Perdew and A. Zunger, Phys. Rev. B {\bf 23}, 5048 (1984). 
\bibitem{BHS}
G. B. Bachelet, D. R. Hamann, and M. Schl\"{u}ter, Phys. Rev. B {\bf 26}, 4199 
(1982).
\bibitem{monk}
H. J. Monkhorst and J. D. Pack, Phys. Rev. B {\bf 13}, 5188 (1976). 
\bibitem{cho}
 J.-H. Cho, Ismail, Z. Y. Zhang, and E. W. Plummer, Phys. Rev. B {\bf59}, 1677 
(1999). 
\bibitem{chen}
S. P. Chen, A. F. Voter, and D. J. Srolovitz, Scripta Metall. {\bf 20}, 1389 
(1986). 
\end{thebibliography}
\end{document}